\documentclass[aps,prl,reprint,groupedaddress]{revtex4-1}
\usepackage{amsmath,amsthm,amssymb,amsfonts}
\usepackage{graphicx,graphics}  
\usepackage{dcolumn}   
\usepackage{setspace}
\usepackage[]{natbib}
\usepackage{color}

\begin{document}

\def\longrightharpoonup{\relbar\joinrel\rightharpoonup}
\def\longleftharpoondown{\leftharpoondown\joinrel\relbar}

\def\longrightleftharpoons{
  \mathop{
    \vcenter{
      \hbox{
	\ooalign{
	  \raise1pt\hbox{$\longrightharpoonup\joinrel$}\crcr
	  \lower1pt\hbox{$\longleftharpoondown\joinrel$}
	}
      }
    }
  }
}

\newcommand{\rates}[2]{\displaystyle
  \mathrel{\longrightleftharpoons^{#1\mathstrut}_{#2}}}


\title{Exact protein distributions for stochastic models of gene expression using partitioning of Poisson processes}


\author{Hodjat Pendar}
\email{hpendar@vt.edu}
\affiliation{Department of Engineering Science and Mechanics,Virginia Tech, Blacksburg, VA 24061}

\author{Thierry Platini} 
\email{thierry.platini@coventry.ac.uk}
\affiliation{Applied Mathematics Research Center, Coventry University, Coventry, CV1 5FB, England}

\author{Rahul V. Kulkarni}
\email{rahul.kulkarni@umb.edu}
\affiliation{Department of Physics, University of Massachusetts, Boston USA}


\date{\today}

\begin{abstract}

Stochasticity in gene expression gives rise to fluctuations in protein
levels across a population of genetically identical cells. Such
fluctuations can lead to phenotypic variation in clonal populations,
hence there is considerable interest in quantifying noise in gene
expression using stochastic models.  However, obtaining exact
analytical results for protein distributions has been an intractable
task for all but the simplest models. Here, we invoke the partitioning
property of Poisson processes to develop a mapping that significantly
simplifies the analysis of stochastic models of gene expression. The
mapping leads to exact protein distributions using results for
mRNA distributions in models with promoter-based regulation. Using
this approach, we derive exact analytical results for steady-state and
time-dependent distributions for the basic 2-stage model of
gene expression.  Furthermore, we show how the mapping leads to 
exact protein distributions for extensions of the basic model that include
the effects of post-transcriptional and post-translational regulation.
The approach developed in this work is widely applicable and can
contribute to a quantitative understanding of stochasticity in gene
expression and its regulation.  
\end{abstract}

\pacs{87.10.Mn, 82.39.Rt, 02.50.-r, 87.17.Aa}

\maketitle

\section{Introduction}

One of the fundamental problems in biology is the elucidation of
molecular mechanisms that give rise to phenotypic variations among
individuals in a population. Recent research has shown that phenotypic
variations can arise without any underlying differences in the
genotype or environmental factors
(\citealp{Collins2011,vanOudenaarden2008}). Such `non-genetic
individuality' is driven by fluctuations (noise) in cellular levels of
gene expression products, as observed in diverse processes ranging
from bacterial persistence (\citealp{Gefen2008}) to HIV-1 viral
infections (\citealp{Weinberger2005}). Quantifying and modeling noise
in gene expression is thus an important step towards a fundamental
understanding of phenotypic variation among genetically identical
cells.

Noise in gene expression is generally analyzed using coarse-grained
stochastic models (\citealp{OShea2004,Paulsson2005}).  For such
models, cellular variations can be characterized using the mean and
variance of mRNA and protein distributions
(\citealp{Paulsson2005,Kondev2008,Coulon2010,Singh2012dynamics}).
However, in several cases, it is of interest to characterize the
entire distribution, rather than just the mean and variance.  For
example, it has been demonstrated that protein distributions can
exhibit features such as bimodality (\citealp{Maheshri2010}) that are
not adequately represented using the first two moments alone. Since
protein levels in single cells can be measured experimentally
(\citealp{Xie2010,Royer2012}), developing analytical approaches for
protein distributions is an important contribution towards building
quantitative models of gene expression.

Given the need for analytical results for the entire distribution,
several approaches have been developed in recent work.  Analytical
results for mRNA distributions have been derived
(\citealp{Peccoud1995,Raj2006,Iyer2009,Zhang2012,Stinchcombe2012,Hornos2005,Hornos2011});
however, the corresponding results for proteins have been
significantly more challenging to obtain. When the mean mRNA lifetimes
($\tau_{m}$) are much shorter than protein lifetimes ($\tau_{p}$),
analytical expressions have been derived for protein steady-state
distributions (\citealp{Friedman2006,Shaherezaei2008}).  More
generally, exact results have recently been derived
(\citealp{Bokes2011}) for the simplest model of gene expression, also
known as the 2-stage model.  While useful results have thus been
obtained, further generalizations are needed to include a broader
class of models that include the effects of cellular regulation.

In this paper, we develop an analytical framework that leads to exact
protein distributions for a wide range of stochastic models of gene
expression. In the following section, we provide brief definitions of
some basic concepts used in the analysis.

%
%


\section{Master equation and generating functions}

Defining the probability distribution $\Phi(X,t)$ to find the system under consideration in a given state $X$ at a time $t$, the corresponding master equation 
is given by 
\begin{equation} 
\partial_t \Phi(X,t)=\sum_Y\left[\Phi(Y,t)w^Y_X-\Phi(X,t)w^X_Y \right],
\end{equation}
where $w^X_Y$ is the rate of transition from $X$ to $Y$.

It is often the case that the state of the system ($X$) is fully
characterized by a set of integers ($\{n_j\}$) such as the number of
mRNA, proteins etc. It follows that the probability distribution
becomes $\Phi(\{n_j\},t)$. The corresponding generating
function $G$ (a function of a set of continuous variable $\{x_j\}$)
is defined by
\begin{equation} 
G(\{x_j\},t)=\sum_{\{n_j\}}x_1^{n_1} x_2^{n_2} ... x_q^{n_q} \Phi(\{n_j\},t).
\end{equation}
All the moments of the probability distribution $\Phi(\{n_j\},t)$ can
be obtained from $G$ by succesive differentiation. Finally, the entire
probability distribution can also be obtained from the expression for
$G$, either analytically or by using numerical approaches. In the
following, we develop an analytical framework for obtaining the
generating function $G$ for protein distributions in stochastic models
of gene expression.

\section{Mapping to reduced models} 
We will consider models of gene expression for which the
creation of mRNAs is a Poisson process occurring with rate
$k_m$. Invoking a well-known theorem on the partitioning of Poisson
processes (\citealp{Ross2006}), we develop a mapping that
significantly simplifies analysis of such models.

\begin{figure}
\hspace{-0.8cm}
\includegraphics[width=.5\textwidth]{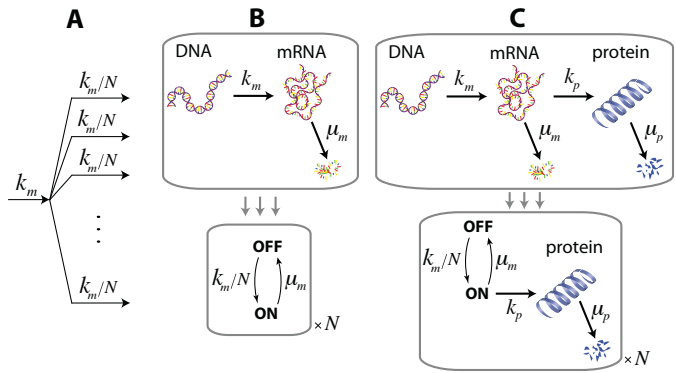}
\caption{(A) A Poisson arrival process with arrival rate $k_m$ can
  be partitioned to $N$ independent and identical Poisson arrival
  processes, each occurring with rate $k_m/N$. (B) Partitioning of the
  Poisson arrival process leads to a mapping from a simple model of
  creation and decay of mRNAs to $N$ independent, identical 2-state
  systems (in the limit $N \to \infty$). The probability of having $m$ mRNAs
in the original model is equivalent to the probability of having $m$ 
two-state systems in the ON state in the reduced model (C) The same mapping 
applied
  to the 2-stage model of gene expression for proteins. Note
  that the reduced model is identical to a model for
  creation and decay of mRNAs with promoter-based regulation.
}\label{Fig1}
\end{figure}

We begin by partitioning the mRNA arrivals into $N$ `types'
(Fig. \ref{Fig1}A). Given a mRNA arrival at any time $t$, the probability that
it is assigned to type $i$ $(i=1\ldots N)$ is $q_{i} =1/N$. Thus each
mRNA is equally likely to be assigned to one of the $N$ types upon
arrival. Denoting by ${\cal N}_{i}(t)$ the number of arrivals of the
$i^{\mathrm{th}}$ type of mRNA by time $t$, it follows from the
theorem of partitioning of Poisson processes (\citealp{Ross2006}),
that the arrival of each type of mRNA is an independent Poisson
process occurring with rate $k_m/N$ (Fig. 1A). In other words, the
${\cal N}_{i}(t)$ $(i=1\ldots N)$ are independent Poisson random
variables with mean $\langle {\cal N}_{i}(t) \rangle = k_{m}t/N$.

The next step consists of taking the limit $N\rightarrow\infty$ and
leads to the definition of the reduced model. For any given time $t$,
in the limit $N \to \infty$, the probability of arrival of more than
one mRNA of any given type can be neglected. It
follows that the random variable describing the number of mRNAs of a
given type is constrained to the value $0$ or $1$. Effectively, after
partitioning of the Poisson arrival process, the mRNA dynamics can be
replaced by the dynamics of a 2-state system.  Thus, at any time $t$,
we have a mapping from the original system to $N$ identical
subsystems. In the limit $N\rightarrow\infty$, each of these
subsystems corresponds to what will be referred to as a `reduced'
model. Further details on the connection between original and reduced
models is provided in Appendix A. In the following, we will refer to
this approach as the PPA (Partitioning of Poisson Arrivals) mapping.

As an illustration, let us consider the number of mRNAs for the simple
model shown in Fig. \ref{Fig1}B. It is readily derived (e.g. using the Master
equation) that the corresponding steady-state distribution is a
Poisson distribution with mean $k_{m}/\mu_{m}$. This result can also
be obtained using the PPA mapping, as illustrated in Fig. \ref{Fig1}B. The
dynamics of the reduced model (a 2-state model) is defined by the
transitions between 0 mRNA(OFF) and 1 mRNA(ON) states driven by the
rates $k_{m}/N$ and $\mu_m$. Therefore, the steady-state generating
function for mRNAs in t he reduced model is given by $g(z)= (1 -
\frac{k_m/N}{\mu_m + k_{m}/N}) + \frac{k_m/N}{\mu_m + k_{m}/N}z$.
Correspondingly, the generating function for the distribution of mRNAs
in the original model is given by
$G(z)=\lim_{N\rightarrow\infty}[g(z)]^N$. This expression reduces to
the generating function of the Poisson distribution with mean
$k_{m}/\mu_m$, thereby recovering the well-known result. An explicit
derivation illustrating this approaching using the Master equation is
provided in Appendix B.

The preceding argument can be generalized to analyze the distribution
of proteins in stochastic models of gene expression. In order to apply
the PPA mapping, we will consider models for which the protein
production from each mRNA proceeds independently.  Let $P(t)$ be the
random variable corresponding to the number of proteins in the system
at time $t$. Partitioning the mRNAs into $N$ `types', we denote by
$p_{i}$ the random variable corresponding to the number of proteins
created by the $i^{\mathrm{th}}$ type of mRNA. Note that, in the limit
$N\rightarrow\infty$, $p_i$ is the random variable corresponding to
the distribution of proteins in the reduced model.  Since each mRNA
contributes independently, the $p_{i}(t)$ are independent, identically
distributed random variables such that $P =\sum_{i=1}^{N} p_{i}$.
Correspondingly, the generating functions for proteins in the original 
($G(z,t)$)
and reduced ($g(z,t)$) models are related by
\begin{equation}
G(z,t) = \lim_{N\rightarrow\infty}[g(z,t)]^{N}.
\end{equation}
Furthermore, it can be shown (Appendix A) that $(g(z,t)-1) \propto k_m t/N$ leading to
\begin{equation}
G(z,t) = \lim_{N\rightarrow\infty}\exp \left[ N \left( g(z,t) - 1\right) \right]. \label{mapping}
\end{equation} 
The significance of the above mapping lies in the fact that it exactly
maps the original problem (obtaining $G(z,t)$) to a reduced problem
(obtaining $g(z,t)$) which is easier to analyze.  The simplification
provided by this mapping derives from the fact that the number of
mRNAs, which is unbounded in the original model, is effectively
replaced by a 2-state system in the reduced model.  

Using Eq.\ref{mapping}, we can readily connect expressions for the mean and
Fano factor of the original model to the corresponding expressions for
the reduced model (Appendix A).  In particular, we show that the Fano
factors for the original and reduced models are identical (in the
limit $N \to \infty$). This is a useful result since it is
generally easier to obtain the Fano factor for the reduced model.

\section{Exact distributions for the 2-stage model} 
We now show how the PPA mapping directly leads to exact results for protein distributions
in the 2-stage model (Fig. \ref{Fig1}C).  The 2-stage model is the simplest
model of stochastic gene expression and has been widely analyzed in
both theoretical and experimental studies.  While exact results for
steady-state distributions have been derived recently
(\citealp{Bokes2011}), the corresponding results for time-dependent
distributions have not been obtained so far.

Using the PPA mapping (Fig. \ref{Fig1}C), we see that the reduced model
(obtained by replacing each type of mRNA by a 2-state system) for
proteins is equivalent to a model for mRNAs with promoter switching.
An explicit derivation of the reduced model, starting from the Master
equation, is provided in Appendix C. The reduced model has been
studied in previous work and analytical results for the corresponding
mRNA distributions have been obtained (\citealp{Peccoud1995,Raj2006}).
Using these results, the generating function for the steady-state
distribution of proteins in the reduced model is given by
\begin{equation}
g^*(z) = \mbox{$_{1}F_{1}$}
\left( \frac{k_m/N}{\mu_p};\,\frac{\mu_m}{\mu_p};\,\frac{k_p}{\mu_p}(z-1) \right).
\end{equation} 
Now, using Eq.\ref{mapping}, we obtain that the protein steady-state
generating function for the 2-stage model is given by
\begin{eqnarray}
&G^*(z) = \\ \label{twostage}
&\lim_{N\rightarrow\infty}
\exp \left\{
N \left[ \mbox{$_{1}F_{1}$}\left( \frac{k_m/N}{\mu_p};\,\frac{\mu_m}{\mu_p};\,\frac{k_p}{\mu_p}(z-1)\right) - 1
\right] \nonumber
\right\}. 
\end{eqnarray}
Equation 6, derived directly from known results, is equivalent
to the exact result derived recently using a different approach
(Appendix C). The concise derivation presented above highlights a
general point: the PPA mapping approach leads to protein distributions
using results for mRNA distributions for models with promoter-based
regulation.

We now apply the PPA mapping to obtain the time-dependent joint
distribution of mRNAs and proteins in the original model (with
generating function $G(y,z,t)$) using the time-dependent distribution
of proteins in the reduced model (with generating function
$g(z,t)$). As noted, the reduced model is equivalent to a model for
mRNAs with promoter-based regulation and the corresponding result for
the time-dependent generating function of the mRNA distribution has
been derived in previous work (\citealp{Iyer2009}). Using this result
to obtain $g(z,t)$, we derive (Appendix C) that the time-dependent
joint generating function of mRNAs and proteins is given by
\begin{eqnarray}
G(y,z,t) & = &\lim_{N\rightarrow\infty}\exp\left\{N\left[g(z,t)+(y-1)\frac{\mu_p}{k_p}\partial_zg(z,t)   \right. \right. \nonumber \\
&  + &  \left. \left.  \frac{y-1}{k_{p}(z-1)}\partial_tg(z,t)-1\right]\right\}. \label{joint}
\end{eqnarray}
Eq \ref{joint} is the most general exact result for the 2-stage model
of gene expression and all the previously derived results can be
obtained from it by taking appropriate limits.

\section{Exact results for extensions of 2-stage model}

\subsection{A Model with multi-step mRNA processing}
\begin{figure}
\hspace{1.2cm}
\includegraphics[width=.48 \textwidth]{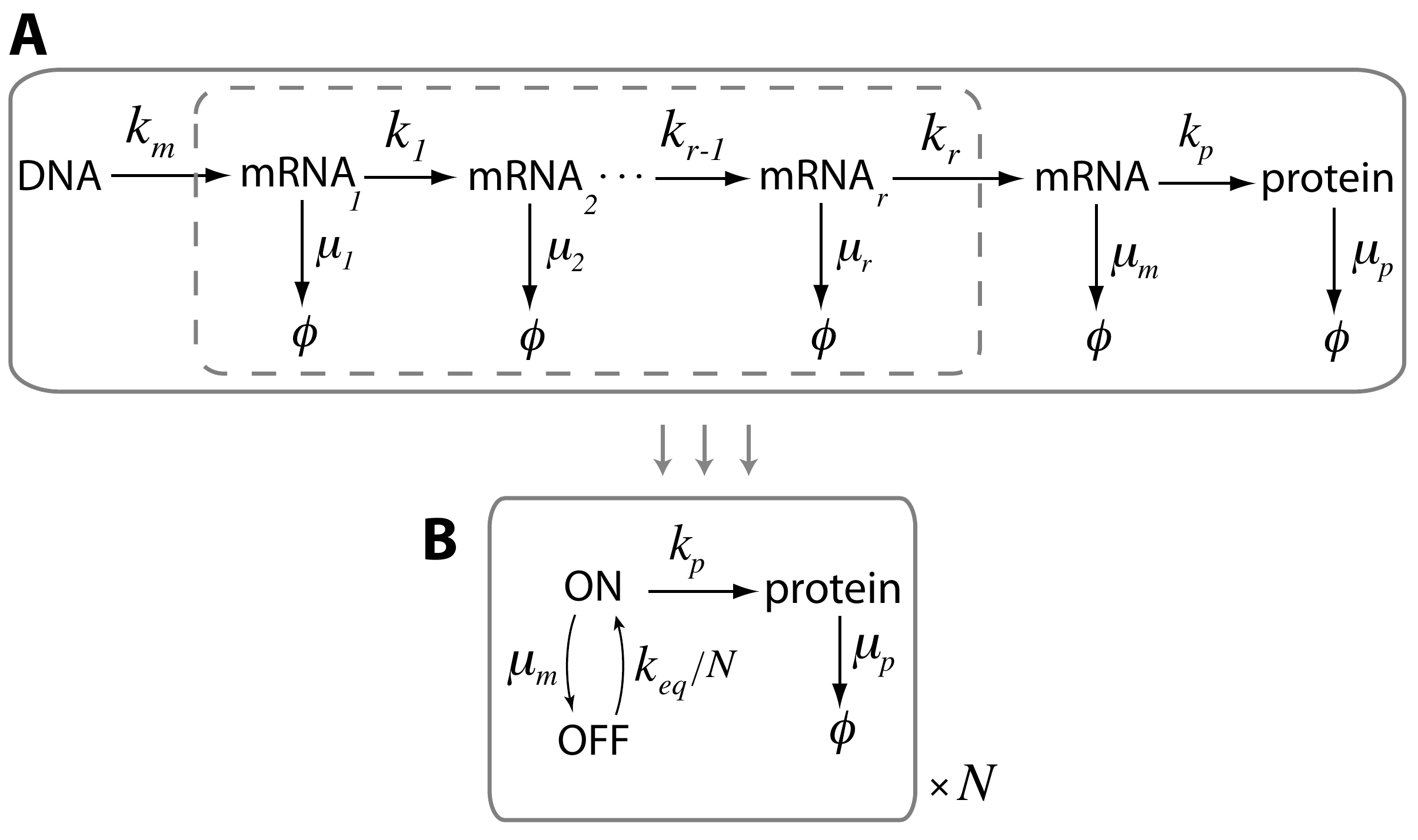}
\caption{(A) In this model mRNAs undergo multi-step pre-processing before being competent to produce proteins. Proteins can be created only from the mature mRNA created in the final processing step. (B)  Arrival of mature mRNAs is shown to be a Poisson process in steady-state leading to the reduced model shown. }\label{Fig2}
\end{figure}

We now show how the partitioning of Poisson processes
leads to exact results for some biologically motivated extensions of
the 2-stage model. Fig \ref{Fig2} presents an extension that allows for an
arbitrary number of processing steps for mRNAs. For example, in
eukaryotes, these processing steps can represent reactions such as
polyadenylation and transport to the cytoplasm which are required for
production of a processed mRNA that is competent for translation. We
will call such a processed mRNA a mature mRNA (whereas the unprocessed
initial transcript will simply be referred to as a mRNA). Let us now
consider the arrival process of a mature mRNA.

The kinetic scheme for the model with $r$ pre-processing steps leading to mature mRNAs is
shown in Fig. \ref{Fig2}A. In the following, we invoke the partitioning
property of Poisson processes to show that the arrival process of a
mature mRNA, in the steady-state limit, is a Poisson process. At any
time $t$, we partition the transcribed mRNAs into 2 types: Type 1
corresponds to a transcribed mRNA that is converted to a mature mRNA
by time $t$ and Type 2 includes all the remaining transcribed
mRNAs. Let us denote the probability that a transcribed mRNA is
classified as Type 1 at time $t$ by $q(t)$. Thus $q = \lim_{t
  \to\infty} q(t)$ is the probability that an mRNA transcribed at
$t=0$ is eventually converted into a mature mRNA. Given an mRNA in the 
$i^{\mathrm{th}}$ state ($1 \le i \le r-1$), the probability that it is converted 
into the $(i+1)^{\mathrm{th}}$ intermediate state without being degraded is
$\left(\frac{k_i}{k_{i} + \mu_{i}} \right)$. Thus, in the long-time limit,
we have
\begin{equation}
 q = \prod_{i=1}^{r} \left(\frac{k_i}{k_{i} + \mu_{i}}\right)
\end{equation}

Note that the arrival process of transcribed mRNAs (Type 1 or Type 2) is 
a Poisson process with rate $k_m$. In the steady-state limit, the probability 
that a transcribed mRNA is labeled as Type 1 is $q$. Thus,  
invoking the partitioning theorem for Poisson processes, we obtain
that the arrival process for a Type 1 mRNA (in the steady-state limit)
is a Poisson process occuring with rate   

\begin{equation}  
k_{eq} = k_{m}\left(\frac{k_{1}}{k_{1} + \mu_{1}}\right) \ldots \left(\frac{k_{r}}{k_{r} + \mu_{r}}\right).
\end{equation}

Since an mRNA is classified at Type 1 once it becomes a mature mRNA,
it follows that the arrival process of mature mRNAs, in the
steady-state limit, is a Poisson process with rate $k_{eq}$. Some
interesting results follow from the preceding observation. First, in
the steady-state limit, since mature mRNAs arrive according to a
Poisson process, the corresponding reduced model is a 2-state model
(as in Fig. \ref{Fig1}B). Thus the steady-state distribution of mature mRNAs is
a Poisson distribution with mean $k_{eq}/\mu_m$.  Furthermore, the
model for proteins is the same as the basic 2-stage model (Fig. \ref{Fig1}C),
but with $k_{m}$ replaced by $k_{eq}$ (Fig. \ref{Fig2}A). Correspondingly, the
exact protein steady-state distribution is given by
Eq. \ref{twostage}, with the substitution $k_m\rightarrow k_{eq}$.
Thus, we obtain that the exact steady-state distribution of proteins
for the model in Fig. \ref{Fig2} is given by
\begin{equation}
G(z) = \lim_{N\rightarrow\infty}
\exp \left\{ 
N \left[ 
\mbox{$_{1}F_{1}$}\left(\frac{k_{eq}/N}{\mu_p};\,\frac{\mu_m}{\mu_p};\,\frac{k_p}{\mu_p}(z-1)\right)-1
\right]
\right\}
\end{equation} \\

\subsection{B Model with delayed degradation }

\begin{figure}
\hspace{-.4 cm}
\includegraphics[width=.38\textwidth]{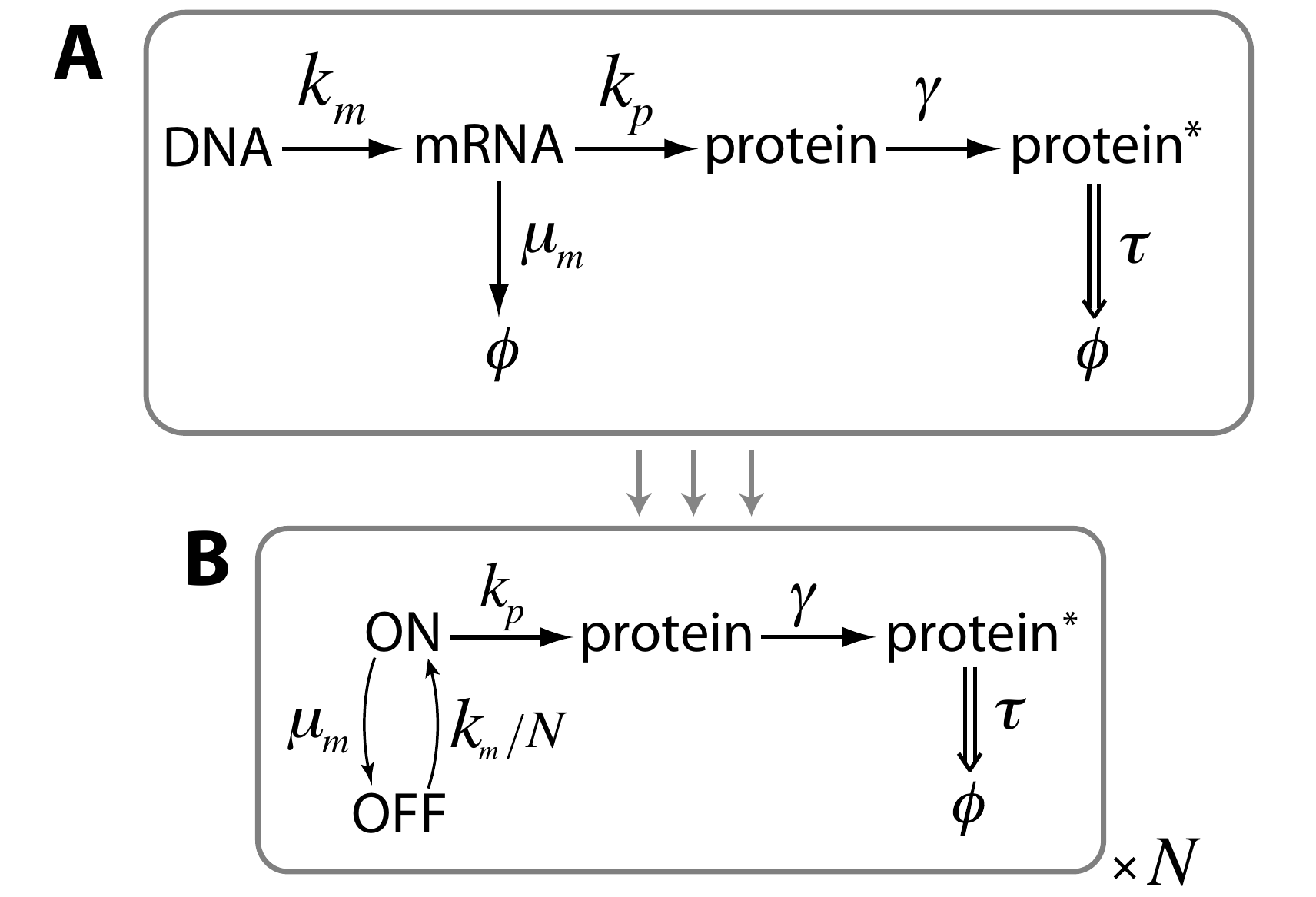}
\caption{(A) Kinetic scheme for model with a fixed-time delay in the degradation of proteins. Protein molecules after being tagged (with rate $\gamma$) are degraded after a fixed time delay $\tau$. (B) Mapping of the original model (A) to  $N$ independent, identical reduced models ($N  \to \infty$) }\label{Fig3}
\end{figure}

The PPA mapping approach can also be applied to models that include
non-Markovian processes. An example involving post-translational
regulation leading to a constant delay in the degradation of proteins
is illustrated in Fig. \ref{Fig3}. The degradation of proteins typically
occurs via complex proteolytic pathways involving multiple steps of
tagging and binding of auxiliary proteins. A simplified assumption
that is commonly used is to replace multi-step degradation by a fixed
time delay, which motivates the model outlined in Fig. \ref{Fig3}. Recent work
has analyzed protein steady-state distributions for models with a
constant time delay in protein degradation
(\citealp{Bratsun2005,Toral2011,Forys2011}). However the processes of
transcription and translation are generally lumped together and it is
assumed that proteins are produced in a single step from the DNA in
these models. The PPA mapping approach allows us to obtain the exact
steady-state protein distributions for a simplified model which
includes both mRNAs and proteins. A detailed derivation (Appendix D) leads to the 
generating function for arbitrary values
of $\tau$. For simplicity, we present here the results in the limit
$\tau\ll1$
\begin{eqnarray} 
&G^*(z) = \exp\left( \frac{k_m k_p \tau (z-1)}{\mu_m-k_p(z-1) } \right)\times   \\
&\lim_{N\rightarrow\infty}\exp\left\{ N \left[ \mbox{$_{1}F_{1}$}\left( \frac{k_m/N}{\gamma};\,\frac{\mu_m}{\gamma};\,\frac{k_p}{\gamma}(z-1)\right) - 1 \right] \right\}. \nonumber
\end{eqnarray}

\section{Discussion}

Several recent experiments have focused on quantifying variations in
gene expression and on inference of the underlying mechanisms based on
observations of noise (\citealp{Munsky2012}). Correspondingly there is
a clear need for theoretical tools to complement such experimental
efforts to understand the role of noise in gene expression in diverse
cellular processes.  The current work addresses this need by
developing an analytical framework for obtaining protein distributions
for stochastic models of gene expression.

We have shown how the partitioning of Poisson arrival processes can lead to
equivalent reduced models that are, in general, simpler to
analyze. This mapping can be used to derive exact results for protein
distributions using mRNA distributions for models with promoter-based
regulation. In recent work, analytical results have been derived for
mRNA distributions for a general class of models with promoter-based
regulation (\citealp{Zhang2012,Stinchcombe2012}).  These results, in
combination with the PPA mapping approach developed in this work, can
be used to obtain exact protein distributions for a broad class of
gene expression models.  Furthermore, previous work
(\citealp{Papoian2006}) has shown how a representation using
generating functions can be used in developing a variational approach
for modeling stochastic cellular processes.  Thus the results obtained
in this work, in combination with such variational approaches, can be
used to provide quantitative insights into the role of different
kinetic schemes in regulating the noise in gene expression.

Noise in gene expression has been shown to play a critical role in
diverse cellular processes (\citealp{Collins2011}).  It is
increasingly becoming clear that quantifying and modeling gene
expression variations among single cells in a population can lead to
fundamental new insights into old problems.  The approach developed in
this work can be used to obtain analytical results for multiple
extensions of the basic gene expression models.
It can be generalized
to analyze models including promoter-based regulation, in particular
the so-called standard model of gene expression
(\citealp{Larson2011}).  
As more cellular processes are studied using
single-cell approaches, the results obtained can guide analysis and
interpretation of such experiments. 
As currently formulated, the approach cannot be used for models 
with feedback effects (i.e with rates that depend on protein numbers), 
however it is hoped that future work will address this issue building on 
current insights.  It will also be of interest to
extend the PPA mapping approach developed in this work to a broader
range of cellular processes for which stochastic effects are critical.

%
%
%
%


\section*{Acknowledgements}
The authors acknowledge funding support from the NSF through award PHY-0957430. TP acknowledges the support of S. Eubank and the NDSSL group at VBI.

\section{Appendix}

\section{A. Connecting original and reduced models}
In this section we discuss the relations between the generating
functions of the original and reduced models. To begin, we note that
the number of mRNAs ($M$) and proteins ($P$) in the original process
are respectively given by the sum of the number of mRNA ($m$) and
protein ($p$) in the $N$ independent and identical reduced
processes. We define $\Phi_M(P,t)$ ($\phi_m(p,t)$)
as the joint probability to find $M$ ($m$) mRNA and $P$ ($p$)
proteins in the original (reduced) process at time $t$. The probability
distributions of the original and reduced processes are related via 
\begin{eqnarray}
&\Phi_M(P,t)=\\
&\sum_{m_i,p_i} \delta\left(M-\sum_im_i\right)\delta\left(P-\sum_ip_i\right) \prod_{i=0}^N  \phi_{m_i}(p_i,t)\nonumber
\end{eqnarray}
where $\delta(X-Y)=1$ for $X=Y$ and zero otherwise. It follows that the generating functions, defined by 
\begin{eqnarray}
G(y,z,t)&=&\sum_{M,P}y^Mz^P\Phi_M(P,t)\\
g(y,z,t)&=&\sum_{m,p}y^mz^p\phi_m(p,t)
\end{eqnarray}
are related by
\begin{eqnarray}\label{Mapping_Equation}
G(y,z,t)=[g(y,z,t)]^N
\end{eqnarray}
as expected for sums of independent and identically distributed random
variables. For large $N$ values, successive differentiation shows that the
averages in both models are related via
\begin{eqnarray}
\bar{m}=\frac{\bar{M}}{N}&\ \ &\bar{m^2}=\frac{\bar{M^2}-\bar{M}^2}{N}\\
\bar{p}=\frac{\bar{P}}{N}&\ \ & \bar{p^2}=\frac{\bar{P^2}-\bar{P}^2}{N}
\end{eqnarray}
Correspondingly the Fano factors for the protein distributions are related by: $F_g=F_G-\bar{P}/N$, so that in the limit $N\rightarrow\infty$ $F_g=F_G$, as presented in the main text.

Focussing our attention on the protein distributions, we choose to
write $G(z,t)=G(1,z,t)$ and $g(z,t)=g(1,z,t)$. In the following, we
consider the limit $N\rightarrow\infty$. In this case, upto any
time $t$, the production of more than one mRNA by the reduced process
is highly unlikely (of second order in $k_mt/N$) as shown in the main
text. In the reduced model, one can therefore neglect all states with
more than one mRNA. Thus we have
\begin{eqnarray}
g(y,z,t)=g_0(z,t)+yg_1(z,t)
\end{eqnarray}
with $g_m(z,t)=\sum_{p}z^p\phi_m(p,t)$. It follows that
\begin{eqnarray}
g(y,z,t)=g(z,t)+(y-1)g_1(z,t)
\end{eqnarray}

In the following we show that, at the lowest order,
the generating function is such that $g(z,t)-1\propto k_mt/N$. Let us denote by
$\phi_m(p,t|m',p',s)$
the probability distribution  at time $t$ with the following condition
$\phi_m(p,t=s|m',p',s)=\delta_{m,m'}\delta_{p,p'}$. Since the transition rate from the $0$ mRNA state to the $1$ mRNA state can be made
arbitrarily small ($k_m/N$), we can assume that the system
has, at maximum, one transition from the state $0$ to $1$ (in a given time
$t$). Neglecting all events that include more than one
transition $0\rightarrow1$,  it follows that $\phi(p,t|0,0,0)$ defined
by $\phi_0(p,t|0,0,0)+\phi_1(p,t|0,0,0)$ can be written has
\begin{eqnarray}
\phi(p,t|0,0,0)&=&\delta(p)e^{-tk_m/N} \\
&+&\int_0^tds\frac{k_m}{N}e^{-sk_m/N} \tilde{\phi}(p,t|1,0,s)\nonumber
\end{eqnarray}
where $\exp(-tk_m/N)$ is the probability that we observe no
$0\rightarrow1$ transitions in a time $t$, while
$\exp(-sk_m/N)k_m/Nds$ is the probability of a transition between time
$s$ and $s+ds$. The distribution $\tilde{\phi}(p,t|1,0,s)$ describes
the probability to find $p$ proteins in a process where all
transitions $0\rightarrow1$ are now neglected, and with the condition
$m=1$ and $p=0$ at time $t=s$. The latter distribution $\tilde{\phi}$,
and its generating function $\tilde{g}$, are therefore independent of
the ratio $k_m/N$. It follows that the generating function $g(z,t)$
(in our case $g(z,t)=g(z,t|0,0,0)$) is
\begin{eqnarray}
g(z,t)&=&e^{-(k_m/N)t}\\
&+&\int_0^tds\frac{k_m}{N}e^{-(k_m/N)s}\tilde{g}(z,t|1,0,s)\nonumber
\end{eqnarray}
which at the first order in $k_m/N$ leads to
\begin{eqnarray}
g(z,t)=1+\frac{k_m}{N}\int_0^t ds\left[\tilde{g}(z,t|1,0,s)-1\right]
\end{eqnarray}
Using the fact that $\tilde{g}(z,t|1,0,s)=\tilde{g}(z,t-s|1,0,0)$ and defining the dimensionless variable $\alpha=1-s/t$ we obtain
\begin{eqnarray}
g(z,t)=1+\frac{k_mt}{N}\int_0^1 d\alpha\left[\tilde{g}(z,\alpha t|1,0,0)-1\right]
\end{eqnarray}
and thus $g(z,t) -1 \propto \frac{k_mt}{N}$ as claimed in the main text. \\

\section{B. 2-Stage model of gene expression: mRNA distribution}

In this section, we show how the Partitioning of Poisson Arrivals
(PPA) mapping leads to the distribution of mRNA levels for the 2-stage
model. In section (A), we write down the master equation
and define the associated generating function $G(z,t)$. The mapping is
then introduced in section (B) by defining the generating function
$g(z,t)$ of the reduced model. The time dependent solution of the
reduced process is given in section (C) and finally the full
generating function $G(z,t)$ is given in section (D).

\subsection{A) Master Equation and Generating function}

The master equation for $\Phi_M(t)$, the probability distribution of
mRNAs in the Fig. \ref{Fig4}A, is given by
\begin{eqnarray}\label{Master_Eq_2_Stage_mRNA}
\partial_t \Phi_M(t)&=&k_m[\Phi_{M-1}(t)-\Phi_M(t)]\\
&+&\mu_m [(M+1)\Phi_{M+1}(t)-M\Phi_M(t)]\nonumber
\end{eqnarray}
The equation for the generating function $G(z,t)=\sum_{M}z^M\Phi_M(t)$ is
\begin{eqnarray}\label{Master_Eq_2_Stage_mRNA_for_G}
\partial_t G&=&k_m(z-1)G-\mu_m (z-1) \partial_zG
\end{eqnarray}
The exact solution can be obtained by directly solving Eq. \ref{Master_Eq_2_Stage_mRNA_for_G}. However, this problem also provides an ideal example to illustrate the PPA mapping approach.
\subsection{B) Mapping}

\begin{figure}
\hspace{1.4cm}
\includegraphics[width=.4\textwidth]{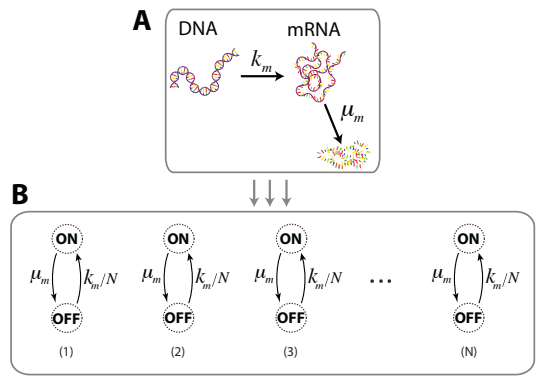}
\caption{(A) The kinetic scheme for a simple model for mRNA production 
and decay. B) Reduced model emerging from the PPA mapping. Probability distribution of number of mRNAs in $(A)$ is identical to the probability distribution of the number of systems in the ON state in $(B)$.}\label{Fig4}
\end{figure}

The PPA mapping connects the original model to $N$ independent, identical reduced models (Fig \ref{Fig4}B).
To explicitly derive it from the Master equation, 
let us write the generating function as $G=(g)^N$. Substituting in
Eq. \ref{Master_Eq_2_Stage_mRNA_for_G}, we see that $g$ and $G$ obey
the same equation with the rescaling $k_m\rightarrow k_m/N$
\begin{eqnarray}\label{Master_Eq_2_Stage_mRNA_for_F}
\partial_t g&=&\frac{k_m}{N}(z-1)g-\mu_m (z-1) \partial_zg
\end{eqnarray}

For the reduced model, defining $\phi_m(t)$ as the probability to have
$m$ mRNAs at time $t$, we can write the generating function as
$g(z,t)= \phi_0(t)+z \phi_1(t)+z^2\phi_2(t)...$. As discussed, for
large $N$, it is unlikely to find more than one mRNA in the reduced
model.  In the stationary state, we have $\phi^*_0 \simeq 1- {\cal
  O}\left({1}/{N}\right)$ and $\phi^*_m \simeq {\cal
  O}\left({1}/{N^m}\right)$ for $m\ge1$.  Keeping the first order term
in $1/N$, the dynamics of the reduced model is effectively described by
the kinetic scheme of an ON-OFF model presented in Fig. \ref{Fig4}B.


\subsection{C) The reduced model: its time dependent solution}

Let us now consider the initial condition $\phi_m(t=0)=\delta_{m,0}$
so that we have $\phi_m(t)\simeq {\cal
  O}\left({1}/{N^m}\right)$ for $m\ge1$ and all time $t$. To first
order in $1/N$, the generating function of the reduced model is 
$g(z,t)=\phi_0(t)+z\phi_1(t)$, where $\phi_0(t)$ and $\phi_1(t)$ obey
the master equation of the 2-state model
\begin{eqnarray}\label{on_off_master_eq}
\partial_t\phi_0(t)=-\partial_t\phi_1(t)=-\frac{k_m}{N}\phi_0(t)+\mu_m \phi_1(t)
\end{eqnarray}
with solution
\begin{eqnarray}
\phi_1(t)=1-\phi_0(t)=\left(1-e^{-(\mu_m+k_m/N)t}\right)\phi^*_1
\end{eqnarray}
where $\phi^*_1=(k_m/N)/(\mu_m+k_m/N)$.


\subsection{D) The full generating function}
The full generating function, is given by 
$G=\lim_{N\rightarrow\infty}(g)^N = \lim_{N\rightarrow\infty}\exp\left[N (g - 1) \right]$ and leads to
\begin{eqnarray}
G(z,t)=\exp\left[ \frac{k_m}{\mu_m}(z-1)(1-e^{-\mu_mt}) \right]
\end{eqnarray}
which corresponds to the well know Poisson distribution of mRNA, with mean $({k_m}/{\mu_m})(1-e^{-\mu_mt})$. \\

\section{C. 2-Stage model of gene expression: protein distribution}

\begin{figure}
\hspace{.5cm}
\includegraphics[width=.48\textwidth]{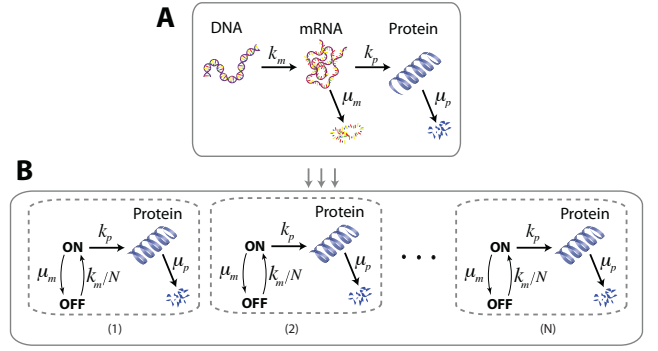}
\caption{(A) The kinetic scheme for protein production in the 2-stage model. (B) Reduced model emerging from the PPA mapping.}\label{Fig5}
\end{figure}

In this section we show how the PPA mapping allows us to obtain the
protein distribution and the joint mRNA-protein distribution for the
2-Stage model (Fig. \ref{Fig5}A). In section (A), we write down the master equation and
define the associated generating function $G(y,z,t)$. Details of the
mapping are presented in section (B) by defining the generating
function $g(y,z,t)$ of the reduced model. The time dependent solution
of $g(y,z,t)$ is given in section (C) and finally the full
generating function $G(y,z,t)$ is obtained in section (D).


\subsection{A) Master Equation and Generating function}

Let us now consider the full probability distribution of the 2-stage
model by writing $\Phi_M(P,t)$ the time-dependent probability
distribution with the master equation:
\begin{eqnarray}\label{Master_Eq_2_Stage}
\partial_t \Phi_M(P,t)&=&k_m[\Phi_{M-1}(P,t)-\Phi_M(P,t)]\\
&+&\mu_m [(M+1)\Phi_{M+1}(P,t)-M\Phi_M(P,t)] \nonumber \\
&+&k_pM[\Phi_M(P-1,t)-\Phi_M(P,t)]\nonumber \\
&+&\mu_p [(P+1)\Phi_M(P+1,t)-P\Phi_M(P,t)]\nonumber
\end{eqnarray}
The generating function 
\begin{equation}
G(y,z,t)=\sum_{M,P}y^Mz^P\Phi_M(P,t) 
\end{equation}
obeys
\begin{eqnarray}
\partial_t G&=&k_m(y-1)G-\mu_m (y-1) \partial_yG \\
&+&k_p(z-1)y\partial_yG-\mu_p (z-1) \partial_zG\nonumber
\end{eqnarray}

\subsection{B) Mapping}
Following the steps presented in the previous section, we define $g(y,z,t)$ 
such that $G=(g)^N$. We see that $g$ is governed by
\begin{eqnarray}\label{Master_Eq_2_Stage_for_F}
\partial_t g&=&\frac{k_m}{N}(y-1)g-\mu_m (y-1) \partial_yg\\
&+&k_p(z-1)y\partial_yg-\mu_p (z-1) \partial_zg\nonumber
\end{eqnarray}
Again, we see that $g$ corresponds to the generating function of the
2-stage model under the rescaling $k_m\rightarrow k_m/N$. For large
$N$ values, the production of two or more mRNA in the reduced model is
unlikely and can be neglected. In the limit $N\rightarrow\infty$ the
generating function can be written as $g(y,z,t)=\sum_p
z^p[\phi_0(p,t)+y\phi_1(p,t)]$. Its dynamics is effectively described by
the kinetic scheme presented in Fig. \ref{Fig5}B. Starting with the initial
condition $\phi_m(p,t=0)=\delta_{m,0}\delta_{p,0}$, we have
$\phi_m(p,t)\simeq 1/N^m$ for $m\ge1$ and $\forall t$.

\subsection{C) The reduced model: its time dependent solution}
Let us write $g$ in the form $g(y,z,t)=g_{0}(z,t)+yg_{1}(z,t)$, where $g_{0}(z,t)$ and $g_{1}(z,t)$ are the generating functions defined by $g_{m}(z,t)=\sum_p z^p\phi_m(p,t)$ ($m=0,1$). The latter quantities obey the coupled equations
\begin{eqnarray}
\partial_t g_0&=&-\mu_p(z-1)\partial_z g_0-\frac{k_m}{N} g_0+\mu_m g_1\\
\partial_t g_1&=&-\mu_p(z-1)\partial_z g_1+k_p(z-1)g_1\nonumber\\
&-&\mu_m g_1+\frac{k_m}{N}g_0
\end{eqnarray}
Summing these two equations and writing $g(z,t)=g(1,z,t)$, we get 
\begin{eqnarray}
g_{1}(z,t) =\frac{1}{k_p(z-1)}\partial_t g(z,t)+\frac{\mu_p}{k_p}\partial_zg(z,t)
\end{eqnarray}
which allows us to write $g(y,z,t)$ as
\begin{eqnarray}
g(y,z,t)&=&g(z,t)+(y-1)\frac{\mu_p}{k_p}\partial_zg(z,t)+\frac{(y-1)}{k_p(z-1)}\partial_tg(z,t)\nonumber\\
\end{eqnarray}

Let us first consider the result for protein distributions in the
stationary state. Based on previous work
(\cite{Peccoud1995,Raj2006,Iyer2009}), we obtain the stationary solution
of the reduced model
\begin{eqnarray}
g^*(z,t)=\mbox{$_{1}F_{1}$}\left(\frac{k_m/N}{\mu_p};\frac{\mu_m}{\mu_p};\frac{k_p}{\mu_p}(z-1)\right)
\end{eqnarray}
where $\mbox{$_{1}F_{1}$}$ is the confluent hypergeometric function. Furthermore, the time-dependent solution for the protein distribution in the reduced model has been obtained in previous
 work (\cite{Iyer2009})
\begin{eqnarray}
g(z,t)&=&F_s(t) \ \mbox{$_{1}F_{1}$}\left(\frac{k_m/N}{\mu_p};\,\frac{\mu_m}{\mu_p};\,\frac{k_p}{\mu_p}(z-1)\right)\\
&+&F_{ns}(t) \ \mbox{$_{1}F_{1}$}\left(1-\frac{\mu_m}{\mu_p}; 2-\frac{\mu_m}{\mu_p}; \ \frac{k_p}{\mu_p}(z-1)\right)\nonumber
\end{eqnarray}
with
\begin{eqnarray}
F_s(t)= \mbox{$_{1}F_{1}$}\left(-\frac{k_m/N}{\mu_p}; \,1-\frac{\mu_m}{\mu_p};\,-\frac{k_p}{\mu_p}e^{-\mu_mt}(z-1)\right)\nonumber\\
\end{eqnarray}
\begin{eqnarray}
F_{ns}(t)&=&\frac{ k_m k_p (z-1)}{N \mu_m (\mu_p-\mu_m)} e^{-\mu_m t}
\\
&\times&\mbox{$_{1}F_{1}$}\left(\frac{\mu_m}{\mu_p}; \,1+\frac{\mu_m}{\mu_p};\,-\frac{k_p}{\mu_p}e^{-\mu_mt}(z-1)\right)\nonumber
\end{eqnarray}

\subsection{D) The full generating function}
From $G=(g)^N$, it is readily shown that the original generating function is given by
\begin{eqnarray}
G(y,z,t)=\lim_{N\rightarrow\infty}e^{N{\cal F}[g(z,t)]}
\end{eqnarray}
with
\begin{eqnarray}
{\cal F}[g(z,t)]&=&g(z,t)+(y-1)\frac{\mu_p}{k_p}\partial_zg(z,t)\nonumber\\
&+& \frac{y-1}{k_p(z-1)}\partial_tg(z,t)-1
\end{eqnarray}
and in the steady-state
\begin{eqnarray}
&G^*(y,z)=\\
&\lim_{N\rightarrow\infty}
\exp\left\{N\left[\mbox{$_{1}F_{1}$}\left(\frac{k_m/N}{\mu_p};\frac{\mu_m}{\mu_p};\frac{k_p}{\mu_p}(z-1)\right)-1\right]\right.\nonumber\\ 
&+ \left. (y-1)\frac{k_m}{\mu_m}\mbox{$_{1}F_{1}$}\left(1; 1+\frac{\mu_m}{\mu_p};\frac{k_p}{\mu_p}(z-1)\right)\right\} \nonumber
\end{eqnarray}

In the following, we show that the steady-state distribution derived above
is equivalent to the exact result derived in recent work (\cite{Bokes2011}). 
By the definition of the hypergeometric functions we have
$\frac{d}{dx}\mbox{$_{1}F_{1}$}(\alpha;\beta;\gamma x)=\frac{\alpha}{\beta}\gamma\ \mbox{$_{1}F_{1}$}(\alpha+1;\beta+1;\gamma x)$
or $\mbox{$_{1}F_{1}$}(\alpha;\beta;\gamma x)=1+ \frac{\alpha}{\beta}\gamma \int_{0}^{x}
\mbox{$_{1}F_{1}$}(\alpha+1;\beta+1;\gamma s) \ \mathrm{d} s $. Using this
relation in the preceding equation for $G^*(z)(=G^*(1,z))$, we obtain:

\begin{eqnarray}
&G^*(z)=\\
&\exp \left\{\frac {k_m k_p}{\mu_m \mu_p} \int_{1}^{z}  \mbox{$_{1}F_{1}$}\left(1;\ 1+\frac{\mu_m}{\mu_p}; \ \frac{k_p}{\mu_p}(s-1)\right) \ \mathrm{d} s \right\}\nonumber
\end{eqnarray}
which is exactly the result derived in previous work (\cite{Bokes2011}). \\


 \section{D. Model with delayed degradation}

We consider an extension of the 2-stage model in which the proteins
degrade in two steps. First proteins are tagged (with rate $\gamma$)
and after being tagged they are degraded with a fixed time delay of
$\tau$ (Fig. \ref{Fig3}A). The corresponding reduced model, obtained using the 
PPA mapping approach, is shown in Fig. \ref{Fig3}B.

To obtain the exact solution for the steady-state protein
distribution, we catergorize the proteins at a given time $t$ (with
$t$ large enough such that the system is in steady-state) into two
groups: tagged and untagged proteins. Then, at time $t + \tau$, all
the tagged proteins will have degraded and the untagged proteins will
survive. During the time-interval $\tau$, mRNAs give rise to new
proteins that are added to the system. These new proteins will also
surive upto time $t + \tau$. Thus, the random variable corresponding
to the number of proteins in the system at time $t + \tau$ is the sum
of two indepedent random variables: the number of untagged proteins at 
time $t$ and the number of proteins created in the time interval 
$[t,t+\tau]$. Let us denote the corresponding generating functions as 
follows: total proteins ($Q(z)$), proteins untagged at time $t$ ($U(z)$)
and proteins created in the time interval $[t,t+\tau]$ ($W(z)$). 
Since the total number of proteins is the sum of the other two independent 
random variables, we have $Q(z)=U(z)W(z)$

The distribution of untagged proteins at time $t$ is the same as the
steady-state distribution of proteins in the basic two-stage model
(with degradation rate in the basic two-stage model set equal to the
tagging rate $\gamma$). The corresponding generating function has been
obtained in previous work {\cite{Peccoud1995}) and is given by
\begin{equation}
U(z)= \lim_{N \to \infty} \mbox{$_{1}F_{1}$}  \left( \frac{k_m}{N\gamma};\ \frac{\mu_m}{\gamma};\ \frac{k_p}{\gamma} (z-1)\right)\ 
\end{equation}
Now, we consider the proteins created in the time interval $\tau$.
For the reduced model, let $W_0(z)$ and $W_1(z)$ be the generating
functions for the protein distribution corresponding to the system
being in OFF and ON states respectively. The following master
equations govern the evolution of $W_0(z)$ and $W_1(z)$ :
\begin{eqnarray}
   \frac{\partial W_0}{\partial t}&=&-\frac{k_m}{N}W_0+\mu_m W_1\\
 \frac{\partial W_1}{\partial t}&=&-\frac{k_m}{N}W_0+\mu_m W_1+k_p(z-1)W_1 
\end{eqnarray}
 therefore:
\begin{eqnarray}
W_1&=&\frac{1}{k_p(z-1)}\frac{\partial  W}{\partial t}\\
W_0&=&\frac{-1}{k_p(z-1)}\frac{\partial W}{\partial t}+W 
\end{eqnarray}
where $W(z)=W_0(z) + W_1(z)$. Correspondingly, we obtain the following equation for $W(z)$:
\begin{equation}
\frac{\partial^2 W}{\partial t^2}+(\frac{k_m}{N}+\mu_m-k_p(z-1))\frac{\partial W}{\partial t}-\frac{k_m}{N}k_p(z-1)W=0
\end{equation}
The solution of this ordinary differential equation is given by {\cite{Peccoud1995}}:
\begin{equation}
W(z,t)=C_1\  e^{ \left(\alpha(z)-\beta(z)\right) t } + C_2\  e^{ \left( \alpha(z)+\beta(z) \right) t } 
\end{equation}
where $\alpha(z)$ and $\beta(z)$ are:
\begin{eqnarray}
2\alpha(z)&=& k_p(z-1)-\mu_m-\frac{k_m}{N}\\
(2\beta(z))^2&=&k_p^2(z-1)^2+2(\frac{k_m}{N}-\mu_m)k_p(z-1)+(\mu_m+\frac{k_m}{N})^2\nonumber\\
\end{eqnarray}
To obtain $C_1$ and $C_2$ we use the initial conditions. Since we are in the steady-state limit, the initial conditions 
are:
\begin{equation}
W_0(z,0)=\frac{\mu_m}{\frac{k_m}{N}+\mu_m}=1-\frac{k_m}{N\mu_m}~~,~~W_1(z,0)=\frac{k_m}{N\mu_m}
\end{equation}
Using the above, we get:
\begin{eqnarray}
C_1&=&\frac{(\beta(z)+\alpha(z))-k_p(z-1)W_1(0)}{2\beta(z)}\\
C_2&=&\frac{(\beta(z)-\alpha(z))+k_p(z-1)W_1(0)}{2\beta(z)}
\end{eqnarray}
For $N \rightarrow \infty$  and $t=\tau$ 
\begin{eqnarray}
W(z,\tau)&=&1+\frac{1}{N} \frac{k_m k_p}{\mu_m^2} \frac{(z-1)}{1-\frac{k_p}{\mu_m}(z-1)} \left[ \mu_m \tau -\frac{k_p}{\mu_m}\right. \nonumber \\
 &\times&\left. \frac{(z-1)}{1-\frac{k_p}{\mu_m}(z-1)} \left( 1-e^{-\mu_m (1-\frac{k_p}{\mu_m} (z-1) ) \tau } \right)   \right]  \nonumber\\
\end{eqnarray}
The generating function of the original model is $G(z)=\lim_{N \to \infty} Q^N$:
\begin{eqnarray}
&G(z)=\\
&\exp\left\{ \frac{k_m}{\mu_m }  \frac{k_p (z-1)}{s(z)}  \left(\mu_m \tau-\frac{k_p(z-1)}{s(z)}  ( 1-e^{- s(z) \tau } ) \right) \right\} 
\nonumber\\
&\times\lim_{N\rightarrow\infty}\exp\left\{N \left( \mbox{$_{1}F_{1}$}\left[ \frac{k_m/N}{\gamma};\,\frac{\mu_m}{\gamma};\,\frac{k_p}{\gamma}(z-1)\right] - 1
\right)
 \right\} \nonumber
\end{eqnarray}
where $s(z)=\mu_m-k_p(z-1)$. \\

\end{document}